\documentclass{article}
\usepackage{amsmath,amsthm}
\usepackage{amssymb,latexsym}
\usepackage[mathscr]{eucal}
\usepackage{setspace}
\usepackage{graphics}
\usepackage{array}

\newcommand{\pr}{\prime}
\newcommand{\lp}{\left(}
\newcommand{\rp}{\right)}
\newcommand{\lb}{\left[}
\newcommand{\rb}{\right]}
\newcommand{\lc}{\left\{}
\newcommand{\rc}{\right\}}
\newcommand{\be}{\begin{equation}}
\newcommand{\ee}{\end{equation}}

\newcommand{\ihat}{\bf\hat{i}}
\newcommand{\nhat}{\bf\hat{n}}
\newcommand{\that}{\bf\hat{t}}

\begin{document}

\title{\textbf{Motion of a Vortex Filament in the Local Induction Approximation: Reformulation of the Da Rios-Betchov Equations in the Extrinsic Filament Coordinate Space}}         
\author{B.K. Shivamoggi\footnote{\normalsize Permanent Address: University of Central Florida, Orlando, FL 32816-1364, USA} ~and G. J. F. van Heijst\\
J. M. Burgers Centre and Fluid Dynamics Laboratory\\
Department of Physics\\
Eindhoven University of Technology\\
NL-5600MB Eindhoven, The Netherlands
}        
\date{}          
\maketitle

\large{\bf Abstract}

In recognition of the highly non-trivial task of computation of the inverse Hasimoto transformation mapping the intrinsic geometric parameter space onto the extrinsic vortex filament coordinate space a reformulation of the Da Rios-Betchov equations in the latter space is given. The nonlinear localized vortex filament structure solution given by the present formulation is in detailed agreement with the Betchov-Hasimoto solution in the small-amplitude limit and is also in qualitative agreement with laboratory experiment observations of helical-twist solitary waves propagating on concentrated vortices in rotating fluids. The present formulation also provides for a discernible effect of the slipping motion of a vortex filament on the vortex evolution.

\pagebreak

\noindent\Large\textbf{1. Introduction}\\

\large The strong nonlinearity of the Euler equations makes the full solution for vortex motion inaccessible so one needs to resort to some mathematical idealizations. In that spirit, Da Rios [1] derived a set of two coupled equations governing the inextensional motion of a vortex filament in an irrotational fluid in terms of time evolution of its intrinsic geometric parameters - curvature and torsion. This derivation, called the local induction approximation (LIA), represents the leading-order behavior of the vortex-induced flow velocity formula once the singularity due to the neglect of the finite vortex core size has been resolved by an asymptotic calculation. The LIA was reinvented by Arms and Hama [2] and the Da Rios equations were rediscovered by Betchov [3] decades later. Hasimoto [4] showed that the Da Rios-Betchov equations can be elegantly combined to give a nonlinear Schrodinger equation. The single-soliton solution of this equation (Zakharov and Shabat [5]) provides a description of an isolated loop of helical twisting motion along the vortex line. However, computation of the inverse Hasimoto transformation mapping the intrinsic geometric parameter space onto the extrinsic filament coordinate space is a highly non-trivial task (Sym [6], Aref and Flinchem [7]). A reformulation of the Da Rios-Betchov equations in the extrinsic vortex filament coordinate space provides a useful alternative approach in this regard, which is the objective of this paper.

\vspace{.3in}

\noindent\Large\textbf{2. Reformulation of the Da Rios-Betchov Equations in the Extrinsic Vortex Filament Coordinate Space}\\

\large The motion of the vortex filament in LIA is governed solely by the local features on the filament so distant parts of the filament need to remain sufficiently separated during the motion. The large-amplitude solutions in LIA are hampered by the violation of this premise as in a self-interaction of the vortex filament. The small-amplitude solutions in LIA do not suffer from this drawback\footnote{\normalsize This development would also provide valid descriptions of long bending waves on an isolated vortex core whose displacement is large compared with the core radius.} (Keener [8], who considered small-amplitude torus knot solutions of the equation of self-induced vortex motion). The extrinsic vortex filament coordinate space formulation developed below considers only small-amplitude vortex motions in the LIA model.

The self-induced velocity of a vortex filament in the LIA is given by (Da Rios [1], Arms and Hama [2])
\be\tag{1}
{\bf v} = \gamma \kappa ~{\that}~ \times ~{\nhat},
\ee
where $\that$ and $\nhat$ are unit tangent and unit normal vectors to the vortex filament, respectively, $\kappa$ is the curvature and $\gamma$ is the strength of the vortex filament.

Consider the vortex filament essentially aligned along the x-axis and assume the deviations from the x-axis to be small (Dmitriyev [9]),
\be\tag{2}
{\bf r} = x {\ihat}_x + y (x,t) {\ihat}_y + z (x,t) {\ihat}_ z.
\ee

We then have,
\be\tag{3}
{\bf v} \equiv \frac{d{\bf r}}{dt} = y_t {\ihat}_y + z_t {\ihat}_z
\ee
\be\tag{4}
{\bf\hat{t}} \equiv \frac{d{\bf r}}{ds} = \frac{d{\bf r}}{dx} \frac{dx}{ds} = \lp {\ihat}_x + y_x {\ihat}_y + z_x {\ihat}_z \rp \frac{dx}{ds}
\ee
where s is the arc length. We are assuming small-amplitude vortex motions, so we have
\be\tag{5}
\frac{dx}{ds} = \lp 1 + y^2_x + z^2_x \rp^{-1/2} \approx 1 - \frac{1}{2} \lp y^2_x + z^2_x \rp.
\ee
Dmitriyev [9] restricts himself to the linear regime and drops the nonlinear terms. We proceed further and include the lowest-order nonlinear terms.

Next,
\be\notag
\kappa {\nhat} \equiv \frac{d{\that}}{ds} = \frac{d{\that}}{dx} \frac{dx}{ds} \approx - \lp y_x y_{xx} + z_x z_{xx} \rp \lb 1 - \frac{1}{2} \lp y^2_x + z^2_x \rp \rb {\ihat}_x
\ee
\be\notag
+ \lb y_{xx} \lc 1 - \frac{1}{2} \lp y^2_x + z^2_x \rp \rc - \lp y^2_x y_{xx} + y_x z_x z_{xx} \rp \rb \lb 1 - \frac{1}{2} \lp y^2_x + z^2_x \rp \rb {\ihat}_y
\ee
\be\tag{6}
+ \lb z_{xx} \lc 1 - \frac{1}{2} \lp y^2_x + z^2_x \rp \rc - \lp z_x y_x y_{xx} + z^2_x z_{xx} \rp \rb \lb 1 - \frac{1}{2} \lp y^2 + z^2_x \rp \rb {\ihat}_z.
\ee

Substituting (4) - (6) into equation (1), we obtain
\be\tag{7a}
y_t = -\gamma z_{xx} + \frac{3\gamma}{2} \lp y^2_x + z^2_x \rp z_{xx}
\ee
\be\tag{7b}
z_t = \gamma y_{xx} - \frac{3\gamma}{2} \lp y^2_x + z^2_x \rp y_{xx}.
\ee

Putting,
\be\tag{8}
\Phi \equiv -(y + iz)
\ee
equations (7a, b) can be combined to yield the cubic nonlinear Schrodinger equation\footnote{It may be mentioned that space curves moving under induction laws different from the vortex-induction law (1), as Lamb [10] pointed out, would be related to solutions of other nonlinear evolution equations.}
\be\tag{9}
\frac{1}{i} \Phi_t = \gamma \Phi_{xx} + \frac{3\gamma}{2} |\Phi_x|^2 \Phi_{xx}
\ee
which corresponds to Hasimoto's [4] development in the vortex filament configuration space for the present vortex filament geometry.

\vspace{.3in}

\noindent\Large\textbf{3. Nonlinear Localized Structures on a Vortex Filament in the LIA}\\

\large Look for a localized stationary solution -
\be\tag{10}
\Phi (x, t) = \frac{\nu}{3^{1/3}} \psi \lp x - u \gamma t \rp e^{i \lp \alpha x - c \gamma t \rp}
\ee
and assume $\psi$ is slowly varying; we then obtain
\be\tag{11}
\psi^{\pr \pr} + i \lp 2 \alpha - u \rp \psi^{\pr} + \lp c - \alpha^2 \rp \psi - \frac{3}{2} \nu \alpha^4 \psi^3 = 0.
\ee

Put,
\be\tag{12}
\alpha = \frac{u}{2} ~, ~\beta \equiv \alpha^2 - c.
\ee
The first relation implies that the velocity of propagation of the structure is twice the torsion, in agreement with Hasimoto's [4] solution. Further, $\beta$ is a measure of the curvature. Equation (11) then becomes
\be\tag{13}
\psi^{\pr \pr} - \beta \psi + \frac{1}{2} \nu \alpha^4 \psi^3 = 0
\ee
which admits a solitary wave solution
\be\tag{14}
\psi = \sqrt\frac{4 \beta}{\nu \alpha^4} ~\text{sech} ~\sqrt{\beta} ~\lp x - u \gamma t \rp
\ee
describing a propagating kink on a vortex filament.

Observe that -

\begin{itemize}
  \item[*] $\psi$ is a single-valued function of $x$, in agreement with Hasimoto's [4] solution in the small curvature limit (which, in Hasimoto's notation, corresponds to $T > 1$, where $T$ in the present notation is $T \equiv \alpha/\sqrt{\beta}$);
  \item[*] this kink and its propagation along the vortex filament are totally due to the torsion ($\alpha \not= 0$), as with Hasimoto's [4] solution;
  \item[*] the filament rotates about the $x$-axis with a velocity $c\gamma$ which, for small torsion, is proportional to the curvature (as measured by $\beta$), as observed originally by Betchov [3];
  \item[*] Betchov [3] - Hasimoto [4] solution\\
\be\notag
\left. 
\begin{matrix}
x = s - 2 ~\text{tanh}~s\\
y = 2 ~\text{sech}~s
\end{matrix}
\right\}
\ee
in the small-curvature limit ($s \ll 1$), namely
\be\notag
\left.
\begin{matrix}
x \sim s\\
y \sim \text{sech}~s
\end{matrix}
\right\}
\ee
is indeed the present solution.
\end{itemize}

The above solution is in qualitative agreement with laboratory experiment observations of helical-twist solitary waves propagating on concentrated vortices in rotating fluids (Hopfinger et al. [11]) for the simple idealized picture given by the LIA (some of the assumptions underlying LIA, like the lack of coupling of the vortex motion to the degrees of freedom of the vortex core, are violated in this experiment). These observations showed that -
\begin{itemize}
  \item[*] the solitary waves travel many times their length without appreciable change in shape;
  \item[*] the propagation speed depends on the strength of the vortex filament;
  \item[*] for fixed vortex filament strength and fixed wave shape, waves travel at a speed inversely proportional to wave amplitude (because the wave propagation speed is proportional to torsion in agreement with (12), and for fixed shape, a wave of smaller amplitude has larger torsion in agreement with (14)).
\end{itemize}

\vspace{.3in}

\noindent\Large\textbf{4. Effect of Slipping Motion}\\

\large Consider a vortex filament experiencing a slipping motion along the filament as is the case in a non-ideal fluid (where viscous diffusion materializes) or in a superfluid (where the vortex lines, which are frozen in the superfluid, slip past the normal fluid (Donnelly [12])). We assume that the slipping motion does not cause changes in shape of the vortex filament in time (Kida [13]).

The velocity of an element of the vortex filament is then given by
\be\tag{15}
{\bf v} = \gamma \kappa {\that} \times {\nhat} + U {\that}
\ee
where U is the slipping speed.
Substituting (3) - (6) into equation (15), we obtain
\be\tag{16a}
y_t = -\gamma z_{xx} + \frac{3\gamma}{2} \lp y^2_x + z^2_x \rp z_{xx} + U y_x \lb 1 - \frac{1}{2} \lp y^2_x + z^2_x \rp \rb
\ee
\be\tag{16b}
z_t = \gamma y_{xx} - \frac{3\gamma}{2} \lp y^2_x + z^2_x \rp y_{xx} + U z_x \lb 1 - \frac{1}{2} \lp y^2_x + z^2_x \rp \rb.
\ee

Making the Galilean transformation
\be\tag{17}
q (x, t) \Rightarrow q (x, \tau) ~, ~\tau \equiv t - x/U
\ee
and introducing (8), we obtain
\be\notag
\Phi_\tau = i \gamma \Phi_{xx} + i \frac{\gamma}{2} |\Phi_x|^2 \Phi_{xx} + \frac{U}{2} |\Phi_x|^2 \Phi_x
\ee

Look for a localized stationary solution
\be\tag{18}
\Phi (x, t) = \nu \psi \lp x - u \gamma \tau \rp e^{i \lp \alpha x - c \gamma \tau \rp}
\ee
and assume again $\psi$ is slowly varying, we then obtain
\be\tag{19}
i \psi^{\pr \pr} + \lp u - 2\alpha \rp \psi^{\pr} + i \lp c - \alpha^2 \rp \psi - \frac{i \nu}{2} \alpha^3 \lp \alpha - \frac{U}{\gamma} \rp \psi^3 = 0.
\ee
Putting again
\be\tag{12}
\alpha = \frac{u}{2} ~, ~c \equiv \alpha^2 - \beta.
\ee
equation (19) becomes
\be\tag{20}
\psi^{\pr \pr} - \beta \psi + \frac{\nu \alpha^3}{2} \lp \alpha - \frac{U}{\gamma} \rp \psi^3 = 0
\ee
which admits a solitary wave solution
\be\tag{21}
\psi = \sqrt{\frac{4 \beta}{\nu \alpha^3 \lp \alpha - U/\gamma \rp}} ~\text{sech} ~\sqrt{\beta} \lp x - u \gamma \tau \rp.
\ee
Observe that the effect of the slipping motion along the vortex filament is in general discernible (unless one does some specialized Galilean transformations) especially for weak or moderately strong vortex filaments.

\vspace{.3in}

\noindent\Large\textbf{5. Discussion}\\

\large In recognition of the highly non-trivial task of computation of the inverse Hasimoto transformation mapping the intrinsic geometric parameter space onto the extrinsic vortex filament coordinate space a reformulation of the Da Rios-Betchov equations in the latter space is in order. The nonlinear localized vortex filament structure solution given by the present formulation is in detailed agreement with the Betchov-Hasimoto solution in the small-amplitude limit and is also in qualitative agreement with laboratory experiment observations (Hopfinger et al. [10]) of helical-twist solitary waves propagating on concentrated vortices in rotating fluids. The present formulation has also been shown to provide for a discernible effect of the slipping motion of a vortex filament in the fluid on the vortex evolution. It may be mentioned that the present formulation also appears to be able to provide a sound basis for a useful exploration of an interesting class of problems in superfluid vortex dynamics, as will be reported elsewhere.

\vspace{.3in}

\noindent\Large\textbf{Acknowledgments}\\

\large BKS would like to thank The Netherlands Organization for Scientific Research (NWO) for the financial support.


\vspace{.3in}

\begin{thebibliography}{xxx}

\bibitem{[1]} L. S. Da Rios: \textit{Rend. Circ. Mat. Palermo} \textbf{22}, 117, (1906).

\bibitem{[2]} R. J. Arms and F. R. Hama: \textit{Phys. Fluids} \textbf{8}, 553, (1965).

\bibitem{[3]} R. Betchov: \textit{J. Fluid Mech.} \textbf{22}, 471, (1965).

\bibitem{[4]} H. Hasimoto: \textit{J. Fluid Mech.} \textbf{51}, 477, (1972).

\bibitem{[5]} V. E. Zakharov and A. B. Shabat: \textit{Sov. Phys. JETP} \textbf{34}, 62, (1972).

\bibitem{[6]} A. Sym: \textit{Fluid Dyn. Res.} \textbf{3}, 151, (1988).

\bibitem{[7]} H. Aref and E. P. Flinchem: \textit{J. Fluid Mech.} \textbf{148}, 477, (1984).

\bibitem{[8]} J. P. Keener: \textit{J. Fluid Mech.} \textbf{211}, 629, (1990).

\bibitem{[9]} V. P. Dmitreyev: \textit{Am. J. Phys.} \textbf{73}, 563, (2005).

\bibitem{[10]} G. L. Lamb: \textit{Phys. Rev. Lett.} \textbf{37}, 235, (1976).

\bibitem{[11]} E. Hopfinger, F. K. Browand and Y. Gagne: \textit{J. Fluid Mech.} \textbf{125}, 505, (1982).

\bibitem{[12]} R. J. Donnelly: \textit{Quantized Vortices in Helium II}, Cambridge Univ. Press, (1991).

\bibitem{[13]} S. Kida: \textit{J. Fluid Mech.} \textbf{112}, 397, (1981).

\end{thebibliography}
\end{document}